\documentstyle[twocolumn,prl,aps]{revtex}

\begin{document}

\title{Millikelvin magnetic relaxation measurements of
$\alpha$-Fe$_{2}$O$_{3}$ antiferromagnetic particles}

\author{E. del Barco, J.M. Hern\'andez, M. Duran and J. Tejada}
\address{Departament de F\'\i sica Fonamental, Universitat de Barcelona\\
Diagonal 647, Barcelona, 08028, Spain\\}

\author{R. D. Zysler and M. Vasquez Mansilla}
\address{Centro At\'omico Bariloche, CNEA, 8400 S. C. de Bariloche, RN, Argentina\\}

\author{D. Fiorani}
\address{ICMAT-CNR, Area della Ricerca di Roma CP10, Monterotondo Stazione. (Rome) Italy\\}
\maketitle

\maketitle

\vspace{2em}
\begin{center}{\bf Abstract}\end{center}

\abstract{In this paper we report magnetic relaxation data for
antiferromagnetic $\alpha$-Fe$_{2}$O$_{3}$ particles of 5 nm mean
diameter in the temperature range 0.1 K to 25 K. The average spin
value of these particles $S \simeq 124$ and the uniaxial
anisotropy constant $D \simeq 1.6\times10^{-2}$ K have been
estimated from the experimental values of the blocking temperature
and anisotropy field. The observed plateau in the magnetic
viscosity from 3 K down to 100 mK agrees with the occurrence of
spin tunneling from the ground state $S_{Z} = S$. However, the
scaling $M$ vs $T\ln(\nu_0 t)$ is broken below 5 K, suggesting the
occurrence of tunneling from excited states below this
temperature.}

\vspace{2em}

The search for candidates to study the quantum oscillations of
spin between opposite orientations is of major interest today for
both basic and applied purposes. There are two areas in which this
is extremely important: the study of the spin quantum coherence in
mesoscopic systems \cite{1,2,3}, and the assessment of magnetic
units as hardware for quantum computation \cite{4,5,6}.

The rate of magnetic relaxation of a single domain particle
associated to thermal fluctuations is $\Gamma = \nu \exp(-U/T)$
where $U$ is the energy barrier and $\nu$ is the attempt
frequency. In the case of ensembles of small particles with a
distribution of volumes, the magnetization depends on time, in the
case of thermal relaxation, only through the combination $T\ln(\nu
t)$. The occurrence of magnetic relaxation in a fine particle
system at temperatures where thermal fluctuations vanish has been
explained in terms of quantum tunneling \cite{7,8}. Most of the
experiments carried out in magnetic systems have been performed at
temperatures above 1 K and using ferro and ferrimagnetic
particulate systems with interaction between particles \cite{7}.
There are also interesting measurements of ferrimagnetic and
ferromagnetic single particles \cite{9,10}. In this paper we show
data of relaxation experiments down to mK for a system of
independent antiferromagnetic particles with a narrow size
distribution.

$\alpha$-Fe$_{2}$O$_{3}$ is an antiferromagnet ($T_{N}$ = 960K)
which undergoes a spin-flip transition at the Morin temperature,
$T_{M}$ = 263K. Below $T_{M}$ it is an uniaxial antiferromagnet
with the spins aligned along the trigonal (111) axis, whereas
above $T_{M}$ is a canted antiferromagnet with the spins
perpendicular to (111), except for a slight canting
(0.13$^{\circ}$) from the basal plane, which results in a small
net magnetic moment. There is, however, another contribution to
the net spin of these particles. This is associated with the
number of non-compensated spins expected from the randomness of
the surface core. The Morin temperature reduces as the particle
size decreases tending to vanish for particles smaller than about
8nm \cite{11}. The antiferromagnetic $\alpha$-Fe$_{2}$O$_{3}$
particles were prepared from precursor FeOOH particles following
the route proposed by Zysler et al. \cite{12}. X-ray powder
diffraction pattern shows the hematite structure corundum type of
the particles. Morphological characterization of the particles was
made by using both a commercial light dispersion equipment before
drying the solution and a 200 keV transmission electron
microscopy. The particles show a platelet shape \cite{12,13,14}.
The size distribution is centered at 5 nm and comprised between 3
and 7 nm. ESR measurements were made at the X-band (9.4 GHz) at
temperatures down to 2 K. No single ion resonance line appears in
the spectrum; that is, our sample is free of paramagnetic
impurities.

Magnetization measurements down to 1.8 K were performed by using a
Quantum Design SQUID magnetometer. The very low temperature
magnetic measurements were carried out by using a top loading
$^{3}$He-$^{4}$He dilution refrigerator (Oxford Kelvinox) which
has incorporated a 5 T superconductor magnet. The sample is inside
the liquid mixture and its temperature may be varied between 50 mK
and 1.2 K. The magnetic moment of the sample is registered using a
superconductor gradiometer by the extraction method. This
gradiometer is coupled through a superconducting transformer to a
Quantum Design dc SQUID which is placed near the 1 K pot. The
temperature of the dc SQUID is kept constant as it is thermally
linked to the 1 K pot. The dc SQUID has also been shielded from
the magnetic field created by the superconductor magnet and the
magnetometer has been calibrated by using pure paramagnetic
samples.

In Figure 1 we show the low field ($H = 300$ Oe) magnetization
measurements down to 1.8K. The zero field cooled magnetisation,
ZFC, is mainly due to the fraction of particles that behave
superparamagnetically at a given $T$, while the field cooled
magnetisation, FC, corresponds to the equilibrium value. The inset
of Figure 1 shows the FC data obtained with the dilution
refrigerator down to the lowest temperature ($T = 100$ mK). The
data for both ZFC and FC above 4 K are in agreement with those
reported by B\o dker et al. \cite{15} from magnetic and M\"{o}ssbauer
measurements performed in the Kelvin regime on particles of 16 nm
average size. The blocking temperature for our particles is,
however, larger than that estimated for the particles of B\o dker
et al. \cite{15} which may be due to the increase of the surface
anisotropy when reducing the size of the particles.

The zero field cooled magnetisation at a given temperature, field
and time is given by \cite{7}
\begin{equation}
M(T,H,t) = \frac{m_0^2 H}{2 T}\int^{V_{B}(T,t)}_{0}dV f(V)
V^{2}\;\;\;
\end{equation}
where $m_0$ is the magnetic moment per unit volume of the material
of the particle, $f(V)$ is the volume distribution of the
particles, $V_{B}(T,t) = \frac{k_BT}{K}\ln(\nu_0 t)$ is the
blocking volume at a given temperature $T$ and time $t$, $\nu_0$
is the attempt frequency and $K$ is the magnetic anisotropy energy
density. At $T > T_{B}$, the average blocking temperature, the
integral of equation (1) becomes constant because the moments of
most of the particles are unblocked. That is, above the blocking
temperature the ZFC magnetisation should follow the $1/T$
superparamagnetic Curie law, as it is experimentally observed.

At $T\ll T_{B}$, the ZFC magnetisation depends on the volume
distribution function because the fraction of superparamagnetic
particles contributing to the magnetic signal decreases when the
temperature decreases. Below 1 K, however, the ZFC magnetisation
increases when temperature decreases, with a $1/T$ dependence down
to the lowest temperature of 100 mK. This behaviour can not be due
to paramagnetic impurities, since their presence should be
detected by EPR measurements, which is not the case. This result
can be explained by equation (1) if the relaxation volume $V_{B}$
does not depend on temperature below 1 K. That is, in presence of
quantum relaxation, the temperature in the definition of $V_{B}$
is replaced by the temperature, $T_{c}$, of the crossover from the
classical to the quantum regime. At $T < T_{c}$, quantum
transitions, independently of the volume distribution, result in
the ZFC curve proportional to $1/T$. It may be concluded,
therefore, that there is a fraction of particles whose magnetic
moments never get blocked due to quantum tunneling effects.

The FC data split from the ZFC data at $T = T_{max}$ as they
correspond to the equilibrium magnetisation at each temperature.
For temperatures lower than 5 K the FC data grow with temperature
following a Curie law until the lowest temperature of 100 mK. The
Curie-Weiss temperature, $\theta_{C}$, deduced from the
extrapolation to zero temperature of the FC data measured below 1K
(see inset in figure 1) is $\theta_{C}\simeq 2$ mK, suggesting a
very weak interaction between the magnetic particles \cite{8}.
Using the temperature variation of the called isothermal remanent
magnetization, $M_{TRM}=2M_{ZFC}-M_{FC}$ \cite{16,17,18} we have
deduced the volume distribution of particles, see dashed lines in
figure 2.

All isothermal magnetisation curves, for $T > T_{B}$, are well
fitted by Boltzmann's statistics and follow a $H/T$ scaling when
considering the random distribution of easy axis and the
temperature variation of the magnetic moment of the particles.
Below $T_{B}$, the $M(H)$ curves show hysteresis. Below 1 K the
cycles close at $H \simeq$ 3 T, which roughly represents the
highest particle anisotropy field $H_{an}$. The continuous
increase of both coercitivity and anisotropy field when reducing
the temperature below the blocking, suggest that the Morin
transition does not take place in these small particles \cite{11}.

Magnetic relaxation measurements were performed down to 100mK. In
order to make easier the comparison of the data obtained above
(Quantum Design SQUID magnetometer) and below (dilution
refrigerator) to 1.6K, we have followed the same procedure in all
the temperature range, from 100mK to 30K. At each temperature, a
high magnetic field, $H =$ 4T, is applied and after one hour it
was switched off. The variation of the total magnetisation with
time was recorded for a few hours. The $\ln(t)$ relaxation was
observed for all temperatures below the blocking temperature.
Figure 2 shows the relaxation data plotted as $M(t)$ vs
$T\ln(\nu_0 t)$. The relaxation data collected above 5 K assemble
nicely into the universal curve expected in the case of purely
thermal relaxation \cite{7,16}. The best fit of this scaling is
obtained using $\nu_0 = 10^{8}$ Hz. Below 5 K, see inset of figure
3, there is a systematic departure from the universal curve,
suggesting that non thermal relaxation phenomena are occurring at
these low temperatures until the lowest temperature $T = 100$ mK.
The derivative, $dM/d[T\ln(t/\tau)]$, of the master curve in the
thermal regime represents the volume/barrier height distribution
(see solid lines in Figure 2). It can be then concluded from both
the relaxation measurements and the low field magnetisation data,
that the peak at 5 nm in the distribution of particle sizes is in
good agreement with the data from electron microscopy and light
scattering. Moreover, we also conclude that we do not have
particles with a size smaller than 2 nm.

The magnetic viscosity, $S$, which is independent from the initial
and final states, is

\begin{equation}
S = \frac{1}{M_0-M_{eq}}\frac{dM}{d\ln t} \propto
\frac{TV_{B}f(V_{B})}{K} \propto T^2\;\;\;,
\end{equation}
and its values have been deduced, see Figure 4, from the $M(t)$
data. As equation (3) reads, the viscosity should go to zero as
temperature decreases if only thermal relaxation is considered. On
the other hand, the viscosity values between 3 K and 100 mK remain
constant suggesting the occurrence of quantum relaxation
phenomena. It has also been deduced from magnetic relaxations at
different fields that the magnetic viscosity monotonically
decreases as a function of the magnetic field, see inset of figure
4, reflecting the existence of a maximum relaxation at zero field.
This could be due to resonant spin tunneling between matching spin
levels \cite{19}.

Note that the distribution function $f(V) \sim V^{-2}$, would
mimic the plateau in the viscosity, but it would result also in a
constant ZFC magnetisation and the preservation of the $M$ vs
$T\ln(\nu_0 t)$ scaling in all the temperature range, in
disagreement with the experimental findings. The existence of
canted spins in a surface layer in a spin-glass-like phase has
also been proposed to explain the low temperature magnetic
properties of some ferri- and antiferromagnetic particles
\cite{20,21,22}. Surface spins have multiple configurations for
any orientation of the core magnetisation, and the distribution of
energy barriers should be $f(E) \sim \frac{1}{E}$. This could
explain the constancy of the viscosity at temperature $T < 3$ K,
but it cannot explain the rest of our low temperature experimental
findings.

Let us discuss our results in the frame of the discrete spin level
structure existing in the two potential wells of the magnetic
anisotropy. The spin Hamiltonian of these nanosized
antiferromagnetic particles with a platelet shape may be written,
as a first approximation, in terms of the dominant uniaxial
anisotropy term, $D S_{z}^{2}$, and the Zeeman term due to the
interaction of the net spin, $S$, of the particles with the
external magnetic field

\begin{equation}
H = - D S_{z}^{2} + H' - g \mu_B S H \;\;\;.
\end{equation}

Due to the size distribution and the non uniform shape of
particles we expect to have a distribution of values for $D$ and
$S$. $H'$ stands for other anisotropy terms. The
symmetry-violating terms in the spin Hamiltonian of equation (4)
inducing tunneling are those associated with the transverse
component of both the magnetic field and magnetic anisotropy. The
values of $D$ and $S$ in equation (4) represent the mean values
for all particles. Taking into account that: a) The average
barrier height, $U \equiv DS^2$, is proportional to the average
blocking temperature, $T_B\simeq$ 12K, $U = T_B \ln (\nu_0 t) =$
248K, where $\nu_0 = 10^8$Hz and $t \simeq$ 10 sec is the
experimental time window, and b) the anisotropy field, $H_{an} = 2
D S \simeq 3$ T, is the field value that eliminates the barrier
height between the two spin orientations, we have estimated $S =
124$ and $D \simeq 1.6\times10^{-2}$ K. Writing the relevant
barrier height as $U = K V$, we find that $K \simeq 8\times10^{5}$
erg/cm$^{3}$.

The temperature, $T_{c}$, of the crossover from quantum to thermal
superparamagnetism \cite{7} may be roughly estimated from $T_{c}
\simeq \mu_B(H_{\|} H_{ex})^{1/2}$, where $H_{\|} \simeq$ 3T and
$H_{ex}$ is the exchange field which may be estimated from the
N\'eel temperature, $T_{N} = 960$ K. We have got $T_{c} \simeq 5$
K.

The first term of equation (4) distributes the spin levels in the
two wells of the magnetic anisotropy separated by the energy
barrier $U$. The spin level, $m = S_{Z}$, contributing to the
magnetic signal at each temperature $T$, and for fields much
smaller than $H_{an}$, satisfies $D (S^{2} - m^{2}) \simeq 20T$,
that is at $T \simeq T_{B}$ the contributing levels are those near
the top of the barrier, $m = 0$, while at $T \ll T_{B}$ only the
ground state $m = S$ contributes to the magnetisation relaxation.
This explains why at temperatures just below the blocking the
relaxation is purely thermal. At lower temperatures, however, the
thermal relaxation above the barriers competes with quantum
tunneling from the excited states. The fact that the scaling $M$
vs $T\ln(\nu_0 t)$ is broken below 5 K should correspond,
therefore, to the occurrence of tunneling effects from the above
mentioned levels. The plateau in the viscosity below 3 K should
reflect the quantum tunneling process from the ground state $S_{Z}
= S$, in agreement with the fact that at these low temperatures
only the level $S_{Z} = S$ is populated. Moreover, the very low
temperature ZFC and FC magnetisation curves obey the $1/T$ Curie
law suggesting that the inverse of the tunneling frequency matches
the experimental window time and the particles are "seen"
superparamagnetically on our resolution time.

In conclusion, we have presented magnetic relaxation data on
antiferromagnetic $\alpha$-Fe$_{2}$O$_{3}$ in the milikelvin
range, for which the most plausible interpretation is the
occurrence of spin tunneling \cite{1,18,19,23,24}. The decrease of
the viscosity when magnetic field increases agrees well with the
discrete level structure in the two wells.


\pagebreak

{\bf FIGURE CAPTIONS}\\

{\bf Figure 1}. ZFC and FC magnetization curves. The inset shows
the linear dependendence of 1/$M_{FC}$ on temperature in the
millikelvin regime. The extrapolation of these data to zero
temperature gives $\theta_c \simeq$ 0 mK.
\\

{\bf Figure 2}. Size distribution of $\alpha$-Fe$_{2}$O$_{3}$
particles deduced from the ZFC and FC curves using equation (2)
(dashed line) and from the $M$ vs. $T \ln(\nu_0 t)$ plot
(continuous line).
\\

{\bf Figure 3}.  Magnetization of $\alpha$-Fe$_{2}$O$_{3}$
particles versus $T \ln(\nu_0 t)$. The best fit has been achieved
using $\nu_0 = 10^8 sec^{-1}$.\\
\\

{\bf Figure 4}. The temperature dependence of the magnetic
viscosity S for the $\alpha$-Fe$_{2}$O$_{3}$ particles. The inset
shows the variation of S with field at $T$ = 3 K.\\

\end{document}